# Readable Racetrack Memory via Ferromagnetically Coupled Chiral Domain Walls

Maokang Shen[1], Yue Zhang[1*], Long You[1], Xiaofei Yang[1]

[1]*School of Optical and Electronic Information, Huazhong University of Science and Technology, Wuhan, 430074, PR China*

* Corresponding author**. Electronic mail:** yue-zhang@hust.edu.cn

**Abstract**

Current-induced motion of domain walls (CIMDW) with interfacial Dzyaloshinskii-Moriya interaction (DMI) in heavy metal (HM)/ferromagnetic (FM) metal multilayers have attracted attention owing to their potential application in novel magnetic memories. In recent years, the CIMDW at ultrahigh speed has been observed in a synthetic antiferromagnetic (SAF) multilayer. However, due to the zero net magnetization, the reading of information from the SAF multilayer is still challenging. In this work, we propose a *readable* racetrack memory consisting of a synthetic ferromagnetic multilayer composed of two FM layers with an interlayer FM coupling. One FM layer had an isotropic DMI, while the other had an anisotropic DMI. This difference of DMIs resulted in the opposite tilting directions of the DW planes in the two layers. This tilting was inhibited by a strong interlayer FM coupling, resulting in an increase in the DW velocity and the reduction of the minimum allowed spacing between two adjacent DWs. In addition, the FM coupling enhanced the stray field, and the stored information could be read conveniently using a conventional reading head. Therefore, our proposal paves a way for the fabrication of a racetrack memory with high reading speed, large storage density, and good readability.

Racetrack memories, based on the current induced motion of domain walls (DWs), have attracted attention owing to their potential application in novel magnetic memories with high storage density, high reading speed, and low dissipation[1-2].

In addition to the conventional DWs in ferromagnetic (FM) nanowires[1], special magnetic microstructures, such as skyrmions and DWs with chirality, have been observed in a heavy metal (HM)/FM metal multilayer with an interfacial Dzyaloshinskii-Moriya interaction (DMI)[3-9]. When a current is injected, the skyrmion or the chiral DW can be driven at a high speed due to the combined action of the spin-orbit torque (SOT) and the DMI [2]. However, the strong DMI also tilts the moving DW plane[10-13], limiting the increase in the memory density and the reading speed.

Very recently, an ultrahigh DW moving speed was observed in a multilayer with a synthetic antiferromagnetic (SAF) structure[2,14-16]. Thanks to strong interlayer antiferromagnetic (AFM) exchange coupling, the tilting of the paired DWs is inhibited[16]. However, the readability of the SAF racetrack is still questionable since the net magnetization of a memory unit (the paired domains) is zero, resulting in a weakly detectable stray field.

In this work, we propose a readable racetrack memory with high reading speed and high storage density in a synthetic ferromagnetic (SF) structure, consisting of two FM layers with an interlayer FM coupling. One FM layer exhibits a conventional isotropic DMI (iDMI), while the other has an anisotropic DMI (aDMI), which was recently discovered in materials with special symmetries[17,18], including a HM/FM bilayer such as W/Co and W/Fe[18,19]. We found that the DWs with iDMI and aDMI tilted toward opposite directions, and the tilting of the paired DWs was inhibited under a strong interlayer exchange coupling. As a result, the DW velocity and the minimum spacing between the neighboring DWs in the SF system were both greater than that in the single FM layer. Additionally, there was also strong stray field of the paired DWs, which improves the readability.

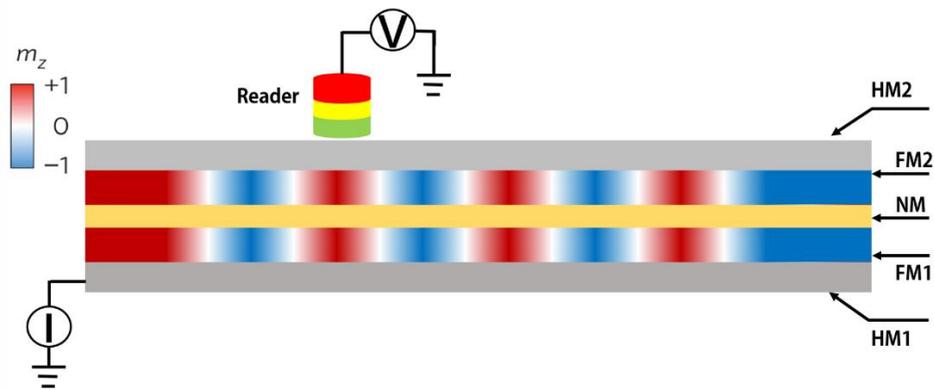

**Fig. 1. Schematic of the model for a readable racetrack memory with an FM coupling. The structure is HM1/FM1/NM/FM2/HM2. Here, the bottom HM1/FM1 has an aDMI, while the top FM2/HM2 exhibits an iDMI. The information can be read using a conventional reader to detect the stray field of the paired DWs.**

Figure 1 shows a schematic of the SF system with an HM1/FM1/NM/FM2/HM2 structure, where

NM denotes normal metal layer acting as a medium for the Ruderman–Kittel–Kasuya–Yosida (RKKY)-type FM exchange coupling between FM1 and FM2. The energy density of a general DMI is expressed as[19,20]

$$\varepsilon_{aDM} = D_x(m_z \frac{\partial m_x}{\partial x} - m_x \frac{\partial m_z}{\partial x}) + D_y(m_z \frac{\partial m_y}{\partial y} - m_y \frac{\partial m_z}{\partial y}),  \quad (1)$$

where $m_x$, $m_y$, and $m_z$ are the components of unit magnetization. The aDMI refers to a tensor with different components for different terms of Lifshitz invariants in Eq. (1), i.e., $D_x \neq D_y$[19]; in the present work, $D_x = -D_y$. The conventional iDMI is represented by a scaling constant ($D$), that is, $D_x = D_y$.

In Fig. 1, HM1/FM1 is the bottom bilayer with an aDMI, and FM2/HM2 is the top bilayer with an iDMI. In the simulation, the DMI tensor of the HM1/FM1 was fixed, while the $D$ of FM2/HM2 was varied in the experimentally achievable range. When current is injected in the $x$-direction, a damping-like SOT originating from the spin Hall effect in the HM layer rotates the moments and drives the DW.

In the multilayer, the $D_x$ of HM1/FM1 and the $D$ of FM2/HM2 need to have the same sign to ensure the same chirality, suitable for the interlayer FM coupling. Under this condition, the spin Hall angles of the two HM layers should be opposite as HM1 is below FM1 and HM2 is above FM2. However, the DW in the FM2 layer can move together with that in the FM1 layer even without an HM2 layer if only the interlayer exchange coupling is sufficiently strong[14]. Certain combinations of HM/FM multilayers, such as Ru/W/Co and Co/Ru[21], Pt/Co and Co/Ir/Pt[9], satisfy the above requirements, and the DWs in the two FM layers move in the same direction.

The investigation was carried out numerically using the micromagnetic simulation software OOMMF[22]. We considered the damping-like SOT, and the code for the DMI was expanded to include the aDMI. We considered a 100-nm-wide and 2000-nm-long nanotrack. The dimension of the unit cell was 2 nm (length) × 1 nm (width) × 0.6 nm (thickness). The thickness of each FM layer was 0.6 nm. The parameters for the ultra-thin Co film with perpendicular magnetic anisotropy (PMA) were as follows[20]: the PMA constant $K = 8 \times 10^5$ J/m$^3$, the saturation magnetization $M_S = 7 \times 10^5$ A/m, the exchange stiffness constant $A = 1 \times 10^{11}$ J/m, and the spin Hall angle $\theta_{SH} = 0.3$ (the $\theta_{SH}$ of W). Further parameter such as the DMI constant was varied between −0.05 mJ/m$^2$ and −2.5 mJ/m$^2$, an experimentally achievable range, by modifying the composition and the thickness of the HM layer[21,22]. The current density ($J$) was in the range of $8.3 \times 10^9$–$1.6 \times 10^{11}$ A/m$^2$.

The static DW structures for both the aDMI ($D_x = -D_y = -1.5$ mJ/m$^2$) and iDMI ($D = -1.5$ mJ/m$^2$) exhibit Néel-type structures and left-hand chirality in the inner part of the track (Figs. 2(a) and 2(c)). However, near the track boundaries, the moments have small projections in the $y$-direction due to the DMI-related boundary conditions[22]. The orientation of the moments near the track boundary for the aDMI is opposite to that for the iDMI (the insets of Figs. 2(a) and 2(c)).

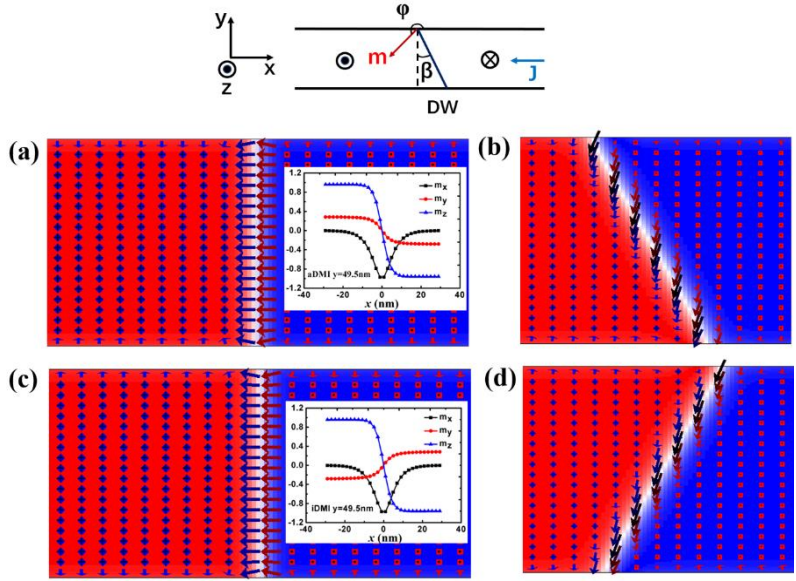

Fig. 2. (a) Structure of the momentum in the static DW with aDMI (the inset shows the variation of $m_x$, $m_y$, and $m_z$ across the DW near the track boundary). (b) Structure of the momentum of a moving DW with aDMI. (c) Structure of the momentum in the static DW with iDMI (the inset shows the variation of $m_x$, $m_y$, and $m_z$ across the DW near the track boundary). (d) Structure of the momentum of a moving DW with iDMI. The upper panel defines the coordinate system, the azimuthal angle of the momentum in the central of the DW, and the tilting angle of the DW plane.

After the static DWs were generated, their motion driven by a current ($J = 1.3 \times 10^{11}$ A/m$^2$) was simulated. The DW tilting can clearly be seen (Figs. 2(b) and 2(d)). However, the tilting orientation of the DW with iDMI is opposite to that for aDMI. In principle, the DMI equals to two effective fields along $x$- and $y$-directions. As $D_x = D_y$ for the iDMI and $D_x = -D_y$ for the aDMI, the $y$ component of the effective field for the iDMI is opposite to that of the aDMI, which can lead to the opposite tilting directions[11]. The SOT-induced motion of the DW with two types of DMI was also analyzed using a cooperative coordinate model (CCM), which is described in detail in the Supplementary Materials (S3).

The SOT-induced motion of the paired DWs in a bilayer track composed of two FM-coupled nanotracks was also investigated ($D_x = -1.5$ mJ/m$^2$ and $J = 1.3 \times 10^{11}$ A/m$^2$), with an interlayer exchange coupling constant of 0.06 mJ/m$^2$. One of the tracks had iDMI, while the other track had aDMI (Track 1). As a comparison, the motion of the DWs in a single track with iDMI (Track 2) or aDMI (Track 3) and that in an FM-coupled (Track 4) bilayer with iDMI were also simulated (Fig. 3).

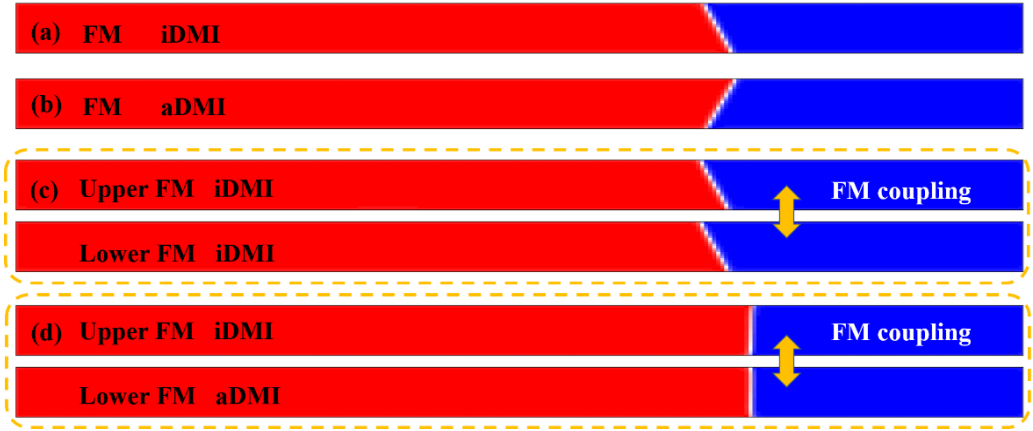

**Fig. 3. SOT-induced motion of (a) a DW with iDMI and (b) a DW with aDMI in a single track, (c) the paired DWs with iDMI in an FM-coupled bilayer track, and (d) the paired DWs in an FM-coupled bilayer track (one layer exhibits iDMI and the other has aDMI). All snapshots were taken when the DWs had been moving for 1 ns. Parameters *J*, *D*, and *σ* are $1.3 \times 10^{11}$ A/m$^2$, $-1.5$ mJ/m$^2$, and 0.06 mJ/m$^2$, respectively.**

In Tracks 2, 3, and 4, the tilting direction for the DW with iDMI is the opposite to that for aDMI. However, in Track 1, the tilting of the paired DW disappears due to a strong interlayer FM coupling. This result is the analogy to the work by Huang et al[20]. They have realized the depression of the skyrmion-Hall effect for the FM coupled skyrmion-antiskyrmion pairs in a similar bilayer. In addition, the velocity of the paired DWs in Track 1 is also obviously higher (the average velocities of the DWs in Tracks 1, 2, 3, and 4 are 460 m/s, 394 m/s, 401 m/s, and 388 m/s, respectively).

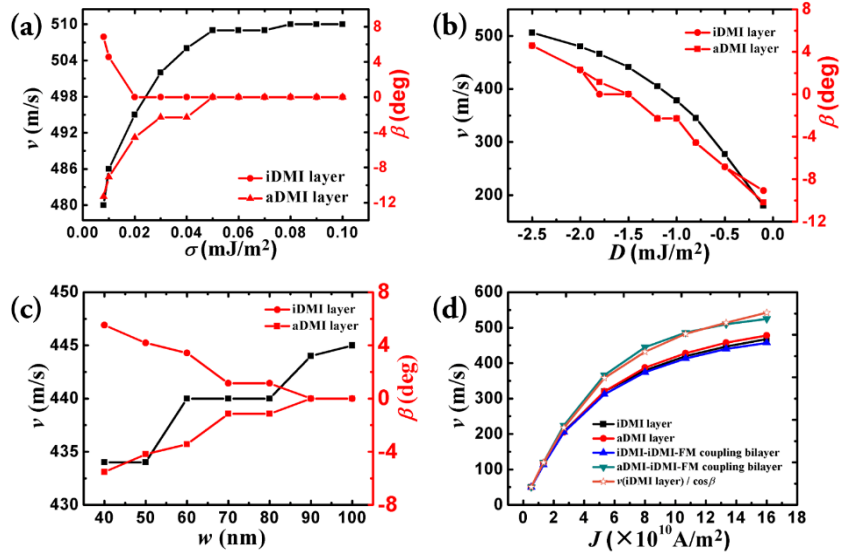

**Fig. 4.** Velocity of the paired DWs and the tilting angles of the DW in the two layers with iDMI and aDMI as a function of (a) the interlayer exchange coupling constant, $\sigma$, (b) the $D$ of the layer with iDMI, (c) the width of the track, $w$. In Fig. (a), $D_x$ of both layers, $J$, and $w$ are fixed as $-1.5$ mJ/m$^2$, $1.3 \times 10^{11}$ A/m$^2$, and 100 nm, respectively. In Fig. (b), $D_x$ of the FM layer with aDMI is fixed as $-1.5$ mJ/m$^2$. $J$, $\sigma$, and $w$ are $8 \times 10^{10}$ A/m$^2$, $-0.06$ mJ/m$^2$, and 100 nm, respectively. In Fig. (c), $D_x$ of both layers, $J$, and $\sigma$ are $-1.5$ mJ/m$^2$, $8 \times 10^{10}$ A/m$^2$, and $-0.06$ mJ/m$^2$, respectively. Figure (d) compares the velocity of the paired DWs as a function of $J$ in different types of tracks. $D_x$ of both layers, $\sigma$, and $w$ are $-1.5$ mJ/m$^2$, $-0.06$ mJ/m$^2$, and 100 nm, respectively. It also indicates that the velocity of the paired DWs in the FM coupled tracks with different DMIs is related to that in the single FM layer by the cosine of the tilting angle.

To determine the optimal parameters for the motion of the DWs in an FM-coupled multilayer, the motion of the paired DWs with different parameters has been investigated in detail (Fig. 4). When the interlayer exchange coupling strength ($\sigma$) increases from 0 to 0.1 mJ/m$^2$, the velocity of the paired DWs increases and the tilting angles of the two FM layers decrease. When $\sigma$ is 0.05 mJ/m$^2$ or larger, the velocity stabilizes, and the tilting of the DWs is completely inhibited.

When the $D$ of the FM layer with iDMI increases from 0 to $-2.5$ mJ/m$^2$ (the $D_x$ for aDMI is fixed at $-1.5$ mJ/m$^2$), the DW velocity keeps increasing, which is accompanied by the rotation of the DW plane. When the absolute value of the $D$ is smaller (greater) than that of the FM layer with aDMI, a negative (positive) tilting angle for the paired DWs is shown. When the $D$ is identical to the $D_x$ of aDMI, the tilting of the paired DWs is completely inhibited.

The track width also affects the velocity and the tilting of the DWs. The velocity of the paired DWs increased with the increase of the width from 40 nm to 100 nm, which is accompanied by the inhibition of the tilting of the DWs in both layers. Near the track boundary, the moment in one layer cannot be parallel with that in the other layer (Fig. 2). As a result, a small tilting of the DW plane near the track boundary is unavoidable. However, this boundary effect becomes negligible when the track width is 90 nm or greater.

The velocity as a function of current density is shown in Fig. 4(d). The DW velocities in the single FM layers with iDMI and aDMI and in the FM-coupled two layers with iDMI are almost the same; however, that in the FM-coupled two layers with different DMIs is higher. Their difference in the velocity keeps increasing with the increase in $J$ and stabilizes (~100 m/s) when $J \geq 8 \times 10^{10}$ A/m$^2$. In addition, it also indicates that the velocity of the tilting DW in the single FM layer can be considered as the projection of the velocity normal to the DW plane in the length direction.

In addition to the reading speed, the storage density is another important characteristic. The storage density is determined by the magnetostatic interaction between the neighboring DWs. However, in the HM/FM bilayer the tilting of the DW plane also restricts the increase of the storage density.

To determine the minimum storage density, an array of domains with different spacing between the neighboring DWs was fabricated in a track. In a single FM layer (Fig. 5(a)), the domain spacing was initially set to 80 nm. After the generation of DWs through relaxation, the domain spacing was reduced to 78 nm. When the DWs were induced to move, the DW planes tilt significantly with an average spacing of approximately 80 nm. When the initial domain spacing was set to be smaller, the neighboring DWs were connected and destroyed when the DWs were induced to move under the same condition. However, in the FM-coupled two layers with distinct DMI (Fig. 5(b)), the allowed minimum initial domain spacing was ~50 nm after the DWs were relaxed and induced to move without tilting. Therefore, the inhibition of the DW tilting by the strong interlayer FM exchange coupling is useful to increase the storage density.

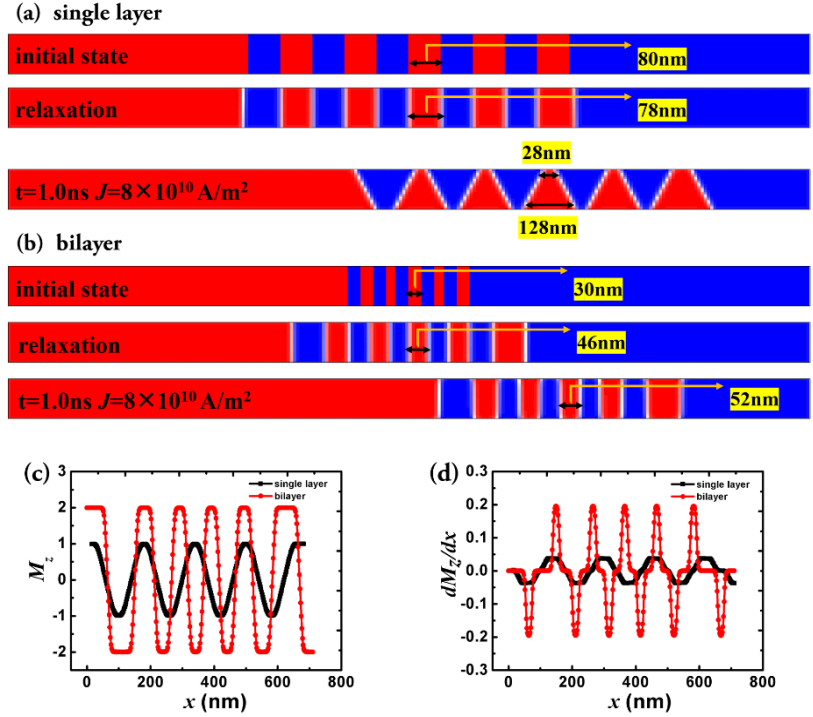

**Fig. 5.** Minimum allowed DW spacing in the single FM layer with iDMI (a) and in the FM coupled two FM layers with different types of DMIs (b). $M_z$ (c) and the differential of $M_z$ (d) with respect to $x$ for a single FM layer (black dots and lines) and for an FM-coupled bilayer (red dots and lines). $D_x$, $\sigma$, and $w$ are −1.5 mJ/m², −0.06 mJ/m², and 100 nm, respectively.

The readability of the information stored in the FM-coupled bilayer is also improved compared to that in the single FM layer due to the enhancement of the stray magnetic field from the double magnetization (Fig. 5(c)). On the other hand, in a single FM layer, the tilting of the DW results in a gradual change of the magnetization from $-M_S$ ($M_S$) of one domain to $M_S$ ($-M_S$) of the adjacent one. However, in the FM-coupled bilayer this transition is abrupt, which reduces the possibility of the error in the reading (Fig. 5(d)).

The motion of the FM-coupled DWs can be analyzed theoretically using the CCM method[11,14]. The equations of motion of the paired DWs with strong interlayer FM coupling can be described as:

$$\frac{2\alpha}{\Delta}\dot{q} + \dot{\varphi}_L + \dot{\varphi}_U = \frac{\pi\gamma_0 H_{SO} J(\cos\varphi_L + \cos\varphi_L)}{2} + 2\gamma_0 H_z \quad (2),$$

$$\frac{\dot{q}}{\Delta} - \alpha\dot{\varphi}_L = \frac{\gamma_0 \pi D \sin(\varphi_L)}{2\Delta\mu_0 M_s} + \frac{\gamma_0 N_x M_s \sin(2\varphi_L)}{2} + \frac{2\sigma\gamma_0 \sin(\varphi_L - \varphi_U)}{\mu_0 M_s t_s} + \frac{\gamma_0 \pi H_x \sin\varphi_L}{2} - \frac{\gamma_0 H_y \pi \cos\varphi_L}{2} \quad (3),$$

$$\frac{\dot{q}}{\Delta} - \alpha\dot{\varphi}_U = \frac{\gamma_0 \pi D \sin(\varphi_U)}{2\Delta\mu_0 M_s} + \frac{\gamma_0 N_x M_s \sin(2\varphi_U)}{2} - \frac{2\sigma\gamma_0 \sin(\varphi_L - \varphi_U)}{\mu_0 M_s t_s} + \frac{\gamma_0 \pi H_x \sin\varphi_U}{2} - \frac{\gamma_0 H_y \pi \cos\varphi_U}{2} \quad (4),$$

where $q$ is the central position of the DWs. The difference in the $q$ of the two layers is neglected [14].

$\varphi_L$ and $\varphi_U$ are the azimuthal angles of the magnetization in the central of the DWs of the lower and the upper layer (Fig. 2). In Eqs. (2)–(4), $\alpha$, $\gamma_0$, $\mu_0$, $N_x$, $t_s$, $H_x$, $H_y$, $H_z$, $H_{SO}$, and $\Delta$ are the Gilbert damping, the gyromagnetic ratio of an electron, the vacuum permeability, the demagnetization factor in the $x$-direction, the thickness of the NM layer, the $x$, $y$, and $z$ components of the external magnetic field, the effective field of the SOT, and the width of domain wall, respectively. $\Delta$, $N_x$, and $H_{SO}$ can be expressed as[12,23]

$$\Delta = \sqrt{A/(K - \mu_0 M_S^2/2)}, \tag{5}$$

$$N_x = L_z \ln 2 / \pi \Delta, \tag{6}$$

and

$$H_{SO} = \mu_B \theta_{SH} J / \gamma_0 e M_S L_z, \tag{7}$$

where $L_z$, $\mu_B$, and e are the thickness of the ferromagnetic layer, the Bohr magneton, and the charge of the electron, respectively.

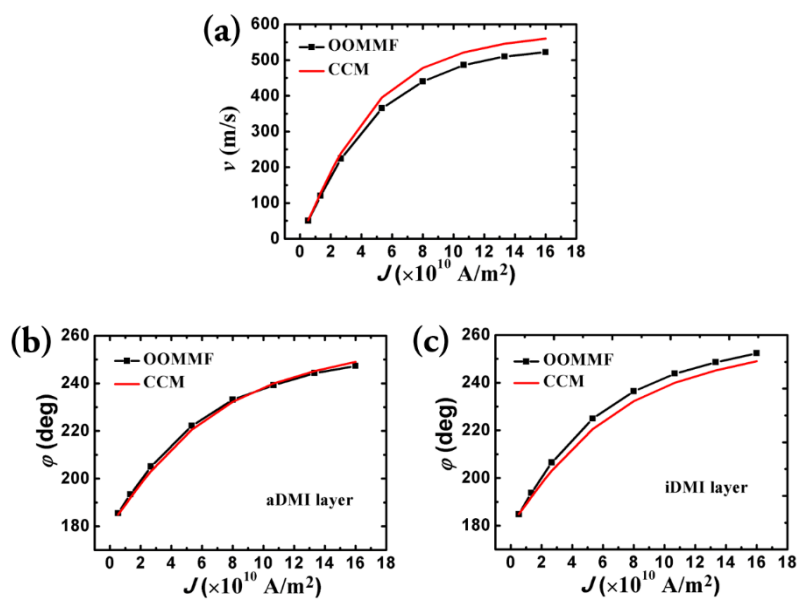

Fig. 6. (a) Velocity of the paired DWs, (b) azimuthal angle of the FM layer with aDMI, and (c) azimuthal angle of the FM layer with iDMI determined by micromagnetic simulation using OOMMF (black dots and lines) and CCM (red line). $D_x$ of both layers, $\sigma$, and $w$ are −1.5 mJ/m², −0.06 mJ/m², and 100 nm, respectively.

Equations (2)–(4) were solved numerically using a fourth order Runge–Kutta algorithm (Fig. 6).

The DW velocity and the azimuthal angles of the DW magnetization in the two FM layers increase with the increasing *J*. The results of the CCM method are in good agreement with those determined by the simulation using OOMMF. The small difference between them can be attributed to the pinning effect near the track boundary that was not considered in the CCM method.

In summary, the SOT-induced DW motion in the FM-coupled HM1/FM1/NM/FM2/HM2 multilayers were investigated numerically. The bottom HM/FM bilayer had aDMI, while the upper bilayer exhibited iDMI. This difference of DMIs resulted in the tilting of the two DWs toward opposite directions. However, when the interlayer exchange coupling was sufficiently strong, this DW tilting could be effectively inhibited. As a result, the DW velocity was increased and the minimum allowed spacing between the neighboring DWs was reduced. These numerical results suggest the possibility of fabricating a racetrack memory with high reading speed, large storage density, and good readability.

**Acknowledgements**

This work was supported by the National Natural Science Foundation of China [grant numbers 11574096, 61674062] and Huazhong University of Science and Technology (No. 2017KFYXJJ037).

**References**

[1]S. S. P. Parkin, M. Hayashi, L. Thomas, Science **320**, 190 (2008).
[2]S.-H. Yang and S. Parkin, J. Phys.: Condens Matter **29**, 303001 (2017).
[3]J. Sampaio, V. Cros, S. Rohart, A. Thiaville, and A. Fert, Nat. Nanotechnol. **8**, 839 (2013).
[4]S. Woo, K. Litzius, B. Krüger, M.-Y. Im, L. Caretta, K. Richter, M. Mann, A. Krone, R. M. Reeve, M. Weigand, P. Agrawal, I. Lemesh, M.-A. Mawass, P. Fischer, M. Kläui, and G. S. D. Beach, Nat. Mater. **15**, 501 (2016).
[5]S. Emori, U. Bauer, S.-M. Ahn, E. Martinez, and G. S. D. Beach, Nat. Mater. **12**, 611 (2013).
[6]K.-S. Ryu, L. Thomas, S.-H. Yang, and S. Parkin, Nat. Nanotech. **8**, 527 (2013).
[7]J. H. Franken, M. Herps, H. J. M. Swagten, and B. Koopmans, Sci. Rep. **4**, 5248 (2014).
[8]S. Emori, E. Martinez, K.-J. Lee, H.-W. Lee, U. Bauer, S.-M. Ahn, P. Agrawal, D. C. Bono, and G. S. D. Beach, Phys. Rev. B **90**, 184427 (2014).
[9]A. Hrabec, N. A. Porter, A. Wells, M. J. Benitez, G. Burnell, S. McVitie, D. McGrouther, T. A. Moore, and C. H. Marrows, Phys. Rev. B **90**, 020402(R) (2014).
[10]K.-S. Ryu, L. Thomas, S.-H. Yang, and S. S. P. Parkin, Appl. Phys. Express **5**, 093006 (2012).
[11]O. Boulle, S. Rohart, L. D. Buda-Prejbeanu, E. Jué, I. M. Miron, S. Pizzini, J. Vogel, G. Gaudin, and A. Thiaville, Phys. Rev. Lett. **111**, 217203 (2013).
[12]E. Martinez, S. Emori, N. Perez, L. Torres, and G. S. D. Beach, J. Appl. Phys. **115**, 213909 (2014).

[13]J. Yun, D. Li, B. Cui, X. Guo, K. Wu, X. Zhang, Y. Wang, J. Mao, Y. Zuo, and L. Xi, J. Phys. D: Appl. Phys. **51**, 155001 (2018).

[14]S.-H. Yang, K.-S. Ryu, and S. Parkin, Nat. Nanotech. **10**, 221 (2015).

[15]Z. Yu, Y. Zhang, Z. Zhang, M. Cheng, Z. Lu, X. Yang, J. Shi, and R. Xiong, Nanotechnology **29**, 175404 (2018).

[16]O. Alejos, V. Raposo, L. Sanchez-Tejerina, R. Tomasello, G. Finocchio, and E. Martinez, J. Appl. Phys. **123**, 013901 (2018).

[17]A. K. Nayak, V. Kumar, T. Ma, P. Werner, E. Pippel, R. Sahoo, F. Damay, U. K. Rößler, C. Felser, and S. S. P. Parkin, Nature **548**, 561 (2017).

[18]M. Hoffmann, B. Zimmermann, G. P. Müller, D. Schürhoff, N. S. Kiselev, C. Melcher, and S. Blüge, Nat. Commun. **8**, 308 (2017).

[19]L. Camosi, S. Rohart, O. Fruchart, S. Pizzini, M. Belmeguenai, Y. Roussigné, A. Stashkevich, S. M. Cherif, L. Ranno, M. Santis, and J. Vogel, Phys. Rev. B **95**, 214422 (2017).

[20]S. Huang, C. Zhou, G. Chen, H. Shen, A. K. Schmid, K. Liu, and Y. Wu, Phys. Rev. B **96**, 144412 (2017).

[21]A. Samardak, A. Kolesnikov, M. Stebliy, L. Chebotkevich, A. Sadovnikov, S. Nikitov, A. Talapatra, J. Mohanty, and A. Ognev, Appl. Phys. Lett. **112**, 192406 (2018).

[22]S. Rohart and A. Thiaville, Phys. Rev. B **88**, 184422 (2013).

[23]Y. Zhang, S. Luo, X. Yang, and C. Yang, Sci. Rep. **7**, 2047 (2017).